\documentclass[aps, prb, twocolumn, showpacs, floatfix,10pt,superscriptaddress]{revtex4-2}
\usepackage{amsmath}
\usepackage{amsfonts}
\usepackage{amsbsy}
\usepackage{amssymb}
\usepackage{graphicx}
\usepackage{textcomp}
\usepackage[caption=false,singlelinecheck=false]{subfig}
\usepackage{xcolor}
\usepackage{mathrsfs}
\usepackage{mathtools}
\usepackage{bm}
\usepackage{gensymb}
\usepackage{braket,wasysym}
\usepackage[colorlinks,linkcolor=magenta,citecolor=red,filecolor=magenta,urlcolor=red,breaklinks]{hyperref}
\usepackage[bottom]{footmisc}
\usepackage{verbatim}

\hypersetup{colorlinks=true, urlcolor=blue, citecolor=magenta, pdfborder={0 0 0}}
\usepackage{breakurl}
\usepackage{natbib}

\allowdisplaybreaks

\begin{document}
\title{\textit{Quantum bridge states} and sub-dimensional transports in quasicrystals 
}

\author{Junmo Jeon}
\email{junmo1996@kaist.ac.kr}
\affiliation{Korea Advanced Institute of Science and  Technology, Daejeon 34141, South Korea}
\author{SungBin Lee}
\email{sungbin@kaist.ac.kr}
\affiliation{Korea Advanced Institute of Science and  Technology, Daejeon 34141, South Korea}

\date{\today}
\begin{abstract}
Exotic tiling patterns of quasicrystals have gotten a lot of attention for unique quantum phenomena such as critical state and multifractality. In this regard, finding new quasi-periodic tiling patterns and the relevant quantum states is one of the main interests in quasicrystals. Here, we focus on new types of quasicrystals described by coexisting phason flips and discuss their quantum transports with distinct local tiling patterns. It turns out that such new tilings are uniquely descended from the translation of high-dimensional lattices and their projection, in particular, a four-dimensional hypercubic lattice. Such unconventional tiling gives rise to sub-dimensionally confined states forming a bridge, so-called \textit{quantum bridge states}. It turns out that electrons in such states could be strongly localized by applying a magnetic field at a particular time but are transported to the sites which share the same local patterns along the bridge with time evolution. 
The speed of transport is controlled via magnetic field strength, and one can even expect electron transportation between multiple bridges where their numbers are determined by the translation vectors of a four-dimensional hypercubic lattice. Our work paves a way to discover anomalous quantum states and their transports uniquely present in quasicrystals with new tiling structures.
\end{abstract}
\maketitle

\section{Introduction}
\label{sec: intro}

Quasicrystal (QC), an ordered but not periodic structure, is famous for exotic tiling patterns with fractality\cite{shechtman1984metallic,PhysRevLett.122.110404,freedman2006wave,kellendonk2015mathematics,aragon2019twisted,GRIMM2003731,van2012formation,baake2017aperiodic,janot2012quasicrystals,suck2013quasicrystals}. They are represented by self-similar structures as a compensation of the absence of translational symmetry\cite{corcovilos2019two,cryst6100124,kawazoe2003structure,suck2013quasicrystals,walter2009crystallography,PhysRevLett.59.1010,baake2017aperiodic,stadnik2012physical,corcovilos2019two}. The tiling patterns of quasicrystals are important because they give rise to  novel quantum states, for instance, a critical wave function, a power-law decaying wave packet, which has been discovered in many distinct quasicrystals\cite{kohmoto1987critical,macia1999physical,PhysRevResearch.3.013168,mace2017critical,deguchi2012quantum,macia1999physical,PhysRevResearch.3.013168,mace2017critical,kohmoto1987critical,fujiwara1988localized,mace2017critical}. Beyond the critical states, another quantum states in quasicrystals have been recently discussed  under magnetic field\cite{PhysRevB.105.045146}. In this case, electrons are strictly localized forming each island with particular length scales, similar to the Aharonov-Bohm-like cage, and such islands are extensively distributed and interfere with each other under field control\cite{vidal1998aharonov,PhysRevB.105.045146}. Such states are uniquely present depending on quasi-periodic pattern. Therefore, discovery of exotic tiling patterns and understanding their quantum phenomena are  important tasks.

In this regard, we focus on quasicrystals with new local tiling patterns, which can be understood by cut-and-project scheme to construct quasicrystalline structures from the high-dimensional lattice\cite{kellendonk2015mathematics,baake2017aperiodic,senechal1996quasicrystals,aragon2019twisted}. 
According to the translations in high-dimensional lattice, the distinct local tiling patterns emerge from the atomic rearrangement,  which is also known as a phason flip\cite{cryst6100124,de2011phonons,fujiwara2007quasicrystals,janssen2004theory,je2021entropic,szallas2009phason,socolar1986phonons}. In particular, it is well known that the Ammann-Beenker tiling results from a four-dimensional (4D) hypercubic lattice and its projection. Interestingly, with particular lattice translations in 4D hypercubic lattice and its projection, the QC structures result in the phason flip and hence the atomic rearrangement gives rise to new local tiling patterns distinct from the well-known Ammann-Beenker tiling\cite{PhysRevLett.59.1010,jeon2021discovery}. In such case, one could ask how electronic wave functions and their transports are characterized.

In this paper, we consider the QC structures with local tiling patterns descended from 4D hypercubic lattice translations,  and study their unique quantum states and transports emerging from new tiling patterns. In detail, we consider two dimensional quasicrystals projected from 4D hypercubic lattice. Distinct from the Ammann-Beenker tiling, certain translations in 4D lattice result in new local tiling patterns forming one dimensional bridges when it is projected. Considering such quasicrystal structures, we demonstrate the sub-dimensionally localized eigenstates forming a bridge itself, so called \textit{a quantum bridge state} in the presence and absence of the magnetic field. Interestingly, under magnetic field, the \textit{quantum bridge states} give rise to an interesting time-flying of the wave function that is not only confined on the \textit{bridges} but also jumping between the sites on the \textit{bridges} which share the same local pattern, leading to the sub-dimensional quantum transport. Moreover, we show that the flux-control could manipulate both the localization of the \textit{quantum bridge states} and the mean current of such sub-dimensional quantum transports. Hence, our finding clarifies the field-controlled sub-dimensional quantum states and dynamics,  purely originated from the novel quasicrystalline tiling patterns.

%

\section{Quantum bridge states}
\label{sec: state}
\begin{figure}[]
\centering
\includegraphics[width=0.5\textwidth]{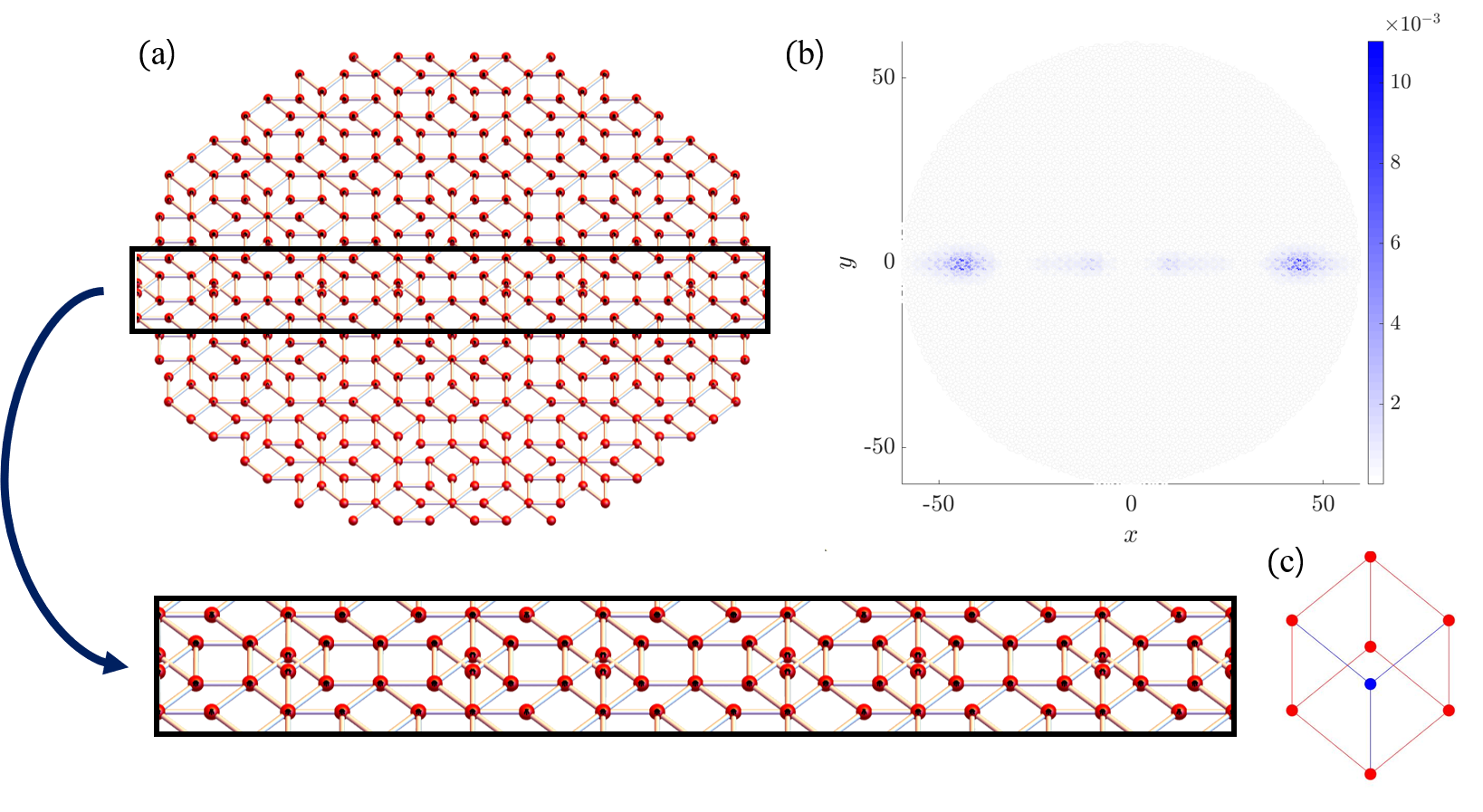}
\caption{\label{fig1} (a) An example of QC descended from the four-dimensional lattice, different from the Ammann-Beenker tiling. It is due to the particular translation, $a(1/2,0,0,0)$, in four-dimensional hypercubic lattice with a lattice constant $a$. The magnified image below shows the unique sub-dimensional local tiling pattern, \textit{bridge} distinct from the Ammann-Beenker tiling. (b) The spatial distribution of a quantum bridge state without magnetic field. The calculation is performed with the number of sites 13694. (c) Typical example of the phason flip from the original atomic arrangement (red color) to the new atomic arrangement (blue color). Outer six atoms are not affected by this phason flip. For the hypercubic lattice translations, blue and red atoms coherently coexist. This gives rise to a distinct sub-dimensional tiling pattern, \textit{bridge} from the Ammann-Beenker tiling. See the main text for details.}
\end{figure}

In this section, we focus on quantum states in new tiling patterns, originated from translations of a 4D hypercubic lattice.
In the cut-and-project scheme, the quasicrystal (QC) is descended from high-dimensional lattice points belonging to the compact subspace\cite{senechal1996quasicrystals,baake2017aperiodic}. Hence, exotic tiling patterns of the QCs depend on the transformation in the high-dimensional lattice\cite{indelicato2012structural,jeon2021discovery}. Particularly, the atomic positions are rearranged by the high-dimensional lattice translations, which is known as a phason flip\cite{de2011phonons,fujiwara2007quasicrystals,janssen2004theory,je2021entropic,szallas2009phason,cryst6100124}. In general, the phason flips do not significantly change the given tiling pattern due to indistinguishable local tiling patterns\cite{fujiwara2007quasicrystals}. However, particular lattice translations in high-dimensional lattice allow an interesting situation where two independent atomic arrangement related by phason flip coexists. For instance, certain translation vectors in four-dimensional hypercubic lattice make the two atoms (colored in blue and red at the center in Fig.\ref{fig1}(c)) coexist, where these two atomic arrangement is related by the phason flips.  In such a way, simple four-dimensional hypercubic lattice shifts could alter the orders of rotational symmetry, and create a distinct sub-dimensional quasiperiodic tiling pattern shown in Fig.\ref{fig1} (a) compared to the eight-fold rotational symmetric Ammann-Beenker tiling\cite{PhysRevLett.59.1010,jeon2021discovery}. Depending on the hypercubic lattice translations, the sub-dimensional tiling pattern as described in a boxed region in Fig.\ref{fig1} (a) appears along at most four different directions\cite{jeon2021discovery}. We name this sub-dimensional quasi-periodic tiling pattern as \textit{bridges}. See Supplemental Material for detailed information of the \textit{bridges} pattern. In Fig.\ref{fig1}(a), we exemplify the QC with a bridge along the $x$-axis resulting from a particular translation vector $a(1/2,0,0,0)$ in 4D hypercubic lattice with a lattice constant $a$.

In the QC shown in Fig.\ref{fig1}(a), let us consider the tight-binding Hamiltonian $H= t \sum_{\left\langle i,j\right\rangle}c_j^{\dagger}c_i +\mbox{h.c.}$
 where $t$ is the uniform hopping magnitude and $i$ and $j$ sites are al length $l=1$ apart from each other, where $l$ is the length of an edge of the rhombus. Since the tiling pattern of the QC in Fig.\ref{fig1} (a) has the local patterns of the Ammann-Beenker tiling except the bridges, 
they share the similar eigenstate characters as the Ammann-Beenker tiling case. The important question that could be asked here is the emergent states forming the sub-dimensional localization on the bridge, so called \textit{quantum bridge states} and the influence of an external fields.

Fig.\ref{fig1} (b) illustrates a quantum bridge state when the magnetic field is absent. Note that the wave function is localized on the bridge along the $x$-axis. 
Let us now consider applying external magnetic field. It turns out that the external magnetic field could enhance the localization due to destructive interference forming Aharonov-Bohm like cages. In detail, the tight-binding Hamiltonian under the magnetic field becomes $H=\sum_{\left\langle i,j\right\rangle}t_{ij}c_j^{\dagger}c_i+\mbox{h.c.}$, where the hopping parameter from the $i$-site to the $j$-site becomes\cite{sakurai2014modern},
\begin{align}
\label{Hamiltonian}
&t_{ij}=t \exp{\left(i\int_{\vec{r}_i}^{\vec{r}_j}\vec{A}(\vec{r})\cdot d\vec{r}\right)}.
\end{align}
Here, $\vec{A}$ is the magnetic vector potential, $\vec{r}_i$ and $\vec{r}_j$ are the position vectors of the $i$ site and the $j$ site,  respectively. Throughout the paper, we choose the Hartree atomic units, $\hbar=e=1$. For Landau gauge, $\vec{A}(\vec{r})=Bx\hat{y}$, where $B$ is the strength of the magnetic field along $z$-axis, let us parametrize the edge from $\vec{r}_{i}=(x_i,y_i)$ to $\vec{r}_{j}=(x_j,y_j)$ as $\vec{r}(s)=(1-s)\vec{r}_{i}+s\vec{r}_{j}$ for $0\le s\le 1$. Then, the hopping parameter with Peierls phase is given by,
\begin{align}
\label{Peierls phase}
&t_{ij}=t \exp{\left(-i\pi\frac{\Phi(x_i+x_j)(y_i-y_j)}{2\phi_0}\right)},
\end{align}
where $\Phi$ is the flux through the unit square and $\phi_0$ is a flux quanta. Then, the Peierls phases in Eq.\eqref{Peierls phase} can give rise to the destructive interference due to the exotic local tiling pattern of QCs, resulting in the enhancement of localization\cite{PhysRevB.105.045146}.
\begin{figure}[]
\centering
\includegraphics[width=0.5\textwidth]{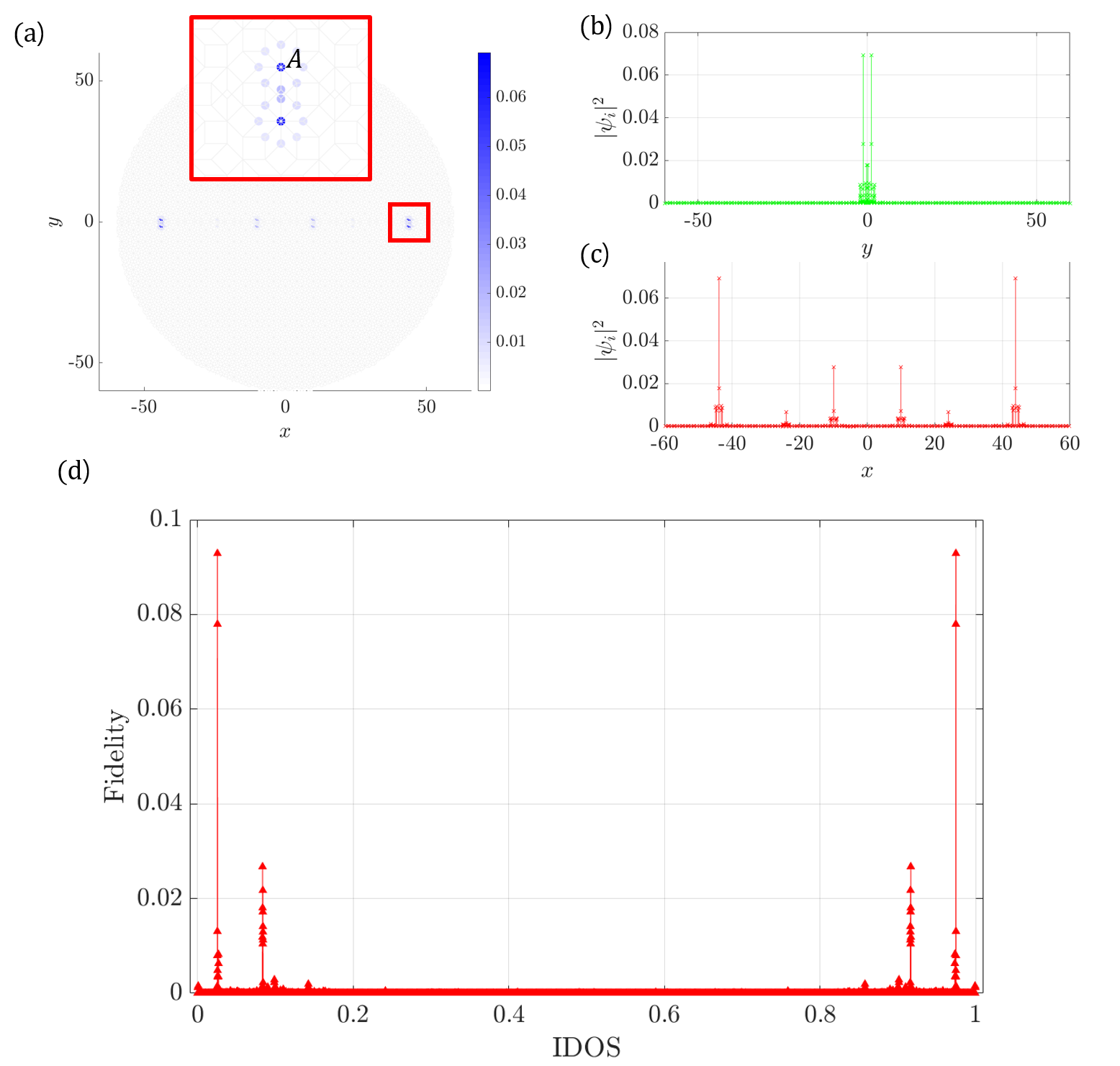}
\caption{
\label{fig: state} 
A quantum bridge state under the magnetic field, $\Phi=1.35\phi_0$. (a) The sub-dimensional quantum bridge state along the $x$-axis. Its probability distributions ($|\psi_i|^2$) are illustrated as functions of the $y$-position (b) or the $x$-position (c). (d) Fidelity of each eigenstate and $\ket{\psi_0}$ as a function of IDOS, where $ \ket{\psi_0}$ is a $\delta$-function-like initial wave function placed on a particular $A$-site in the inset of (a). The quantum bridge state illustrated in (a) has the highest fidelity. Each marker indicates the fidelity with each eigenstate. The calculation is performed with the number of the sites, 13694. See the main text for details.}
\end{figure}
Fig.\ref{fig: state} (a) shows the typical example of the quantum bridge state with $\Phi=1.35\phi_0$. Figs.\ref{fig: state} (b,c) illustrate the probability distributions of the quantum bridge state as the functions of the $y$ and $x$ positions, respectively. Fig.\ref{fig: state} (b) shows that the \textit{quantum bridge state} is more strongly localized by the magnetic field compared to the case without the magnetic field. This is because the Peierls phase in Eq.\eqref{Peierls phase} induces the destructive interference of wave function dependent on the local tiling pattern. 
Furthermore, the probability distribution along the bridge illustrated in Fig.\ref{fig: state} (c) also shows that the wave function is  localized along the bridges under the magnetic field. Specifically, the \textit{quantum bridge state} is highly localized along the sites that share the same local pattern of the $A$-site shown in the inset of Fig.\ref{fig: state} (a). We emphasize that the local pattern of the $A$-site results from  the existence of atoms at both positions related by phason flips and these sites are uniquely emergent due to  the special hypercubic translations. Such sites  are extensively present along the bridge since these sites repeatedly appear along the bridge. 

Now let us consider the time evolution of such localized electron wave function. Let $\ket{\psi_0}$ be the initial state that is the $\delta$-function-like wave function placed on a particular $A$-site as shown in the inset of Fig.\ref{fig: state} (a). For a given set of eigenstates $\{\ket{\phi_n}\}$, where $n$ is the index for $n$-th energy, the initial state would be represented by the linear combination of the eigenstates. Thus, with the coefficients $c_n=\braket{\phi_n|\psi_0}$, 
\begin{align}
\label{decomposition}
&\ket{\psi_0}=\sum_n c_n\ket{\phi_n}.
\end{align}
Its dynamics is given by,
\begin{align}
\label{decomposition}
&\ket{\psi(\tau)}=\sum_n c_n\exp{(-i\tau E_n)}\ket{\phi_n}.
\end{align}
Here, $E_n$ is the $n$-th eigenvalue, and $\tau$ is the time. When almost all of the $c_n$ are negligible except for a few $n$, the dynamics originate from such a few dominant eigenstates. Fig.\ref{fig: state} (d) shows the fidelity, $\left\vert c_n\right\vert^2$ of each $\ket{\phi_n}$ and $\ket{\psi_0}$ as a function of the integrated density of state (IDOS) in the presence of magnetic field($\Phi=1.35\phi_0$). The eigenstates having a significant fidelity are the \textit{quantum bridge states}. 
Thus, under the magnetic field, the time-flying of the state $\ket{\psi_0}$ leads to the sub-dimensional transport along the bridges that is jumping between the sites as we demonstrate in the next section.

\section{Transports along the bridges}
\label{sec: transport}
Now we discuss the sub-dimensional quantum transport on the bridges and its controllability under the magnetic field. Here, the external magnetic field would play a role of the adjuster that controls the transport current. For a concrete argument, we demonstrate the sub-dimensionally confined quantum transports along the bridge on the $x$-axis.

\begin{figure}[]
\centering
\includegraphics[width=0.5\textwidth]{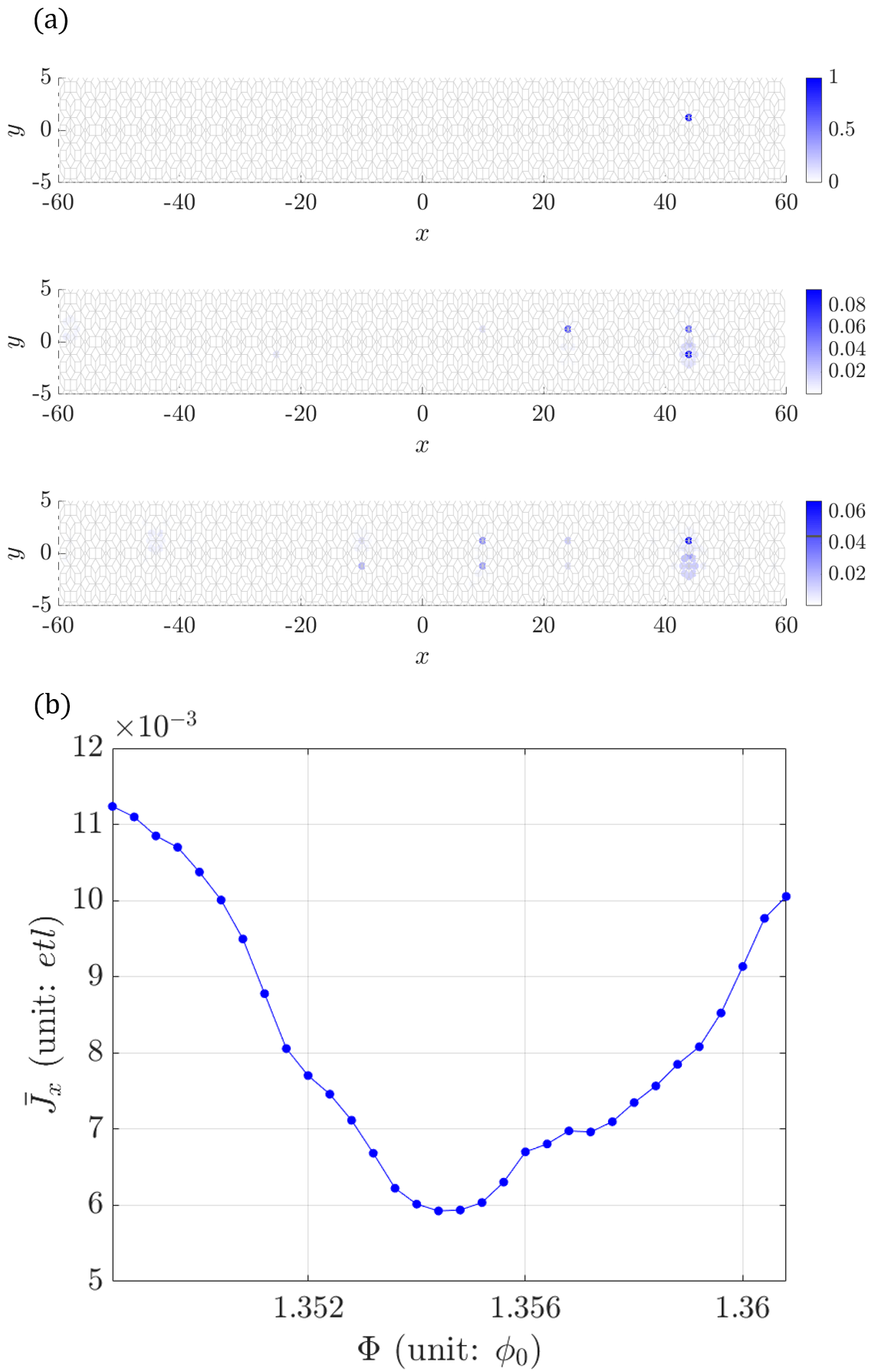}
\caption{\label{fig: transport} Quantum transport demonstrated on a bridge along $x$-axis for $\Phi=1.35\phi_0$. (a) Top panel shows the initial $\delta$-function like wave function at time $\tau=0$. At time $\tau=1500 t^{-1}$ and $3000 t^{-1}$, the electron wave functions evolve to the sites having the same local pattern of the initial site as shown in the middle and bottom panels, respectively. 
 (b) The time-averaged mean current of the transport, $\bar{J}_x$, as a functions of  flux $\Phi$. Current drastically changes with flux. Time-averaged $\bar{J}_x$  is  calculated during the time $\tau=3000 t^{-1}$. See the main text for details.
}
\end{figure}

Initially, let us consider the $\delta$-function-like wave function placed on the $A$-site shown in the inset of Fig.\ref{fig: state} (a). Fig.\ref{fig: transport} (a) illustrates its time-flying. The top panel of Fig.\ref{fig: transport} (a) shows the initial state $\ket{\psi_0}$. With the time evolution $\tau$, the wave function jumps to the sites having the same local pattern on the bridge (See the middle and bottom panels of Fig.\ref{fig: transport} (a) for $\tau=1500 t^{-1}$ and $\tau=3000 t^{-1}$.). This results in sub-dimensional transport along the bridges.

One could ask how such transport could be controlled via magnetic flux change. For quantitative argument, we consider the mean current defined by,
\begin{align}
\label{velocity}
&\bar{J}_x=-e\frac{i}{\tau}\int_0^\tau d\tau'\left\langle [\hat{x},H]\right\rangle(\tau'). 
\end{align}
Here, $\tau$ is the time for averaging. As shown in Fig.\ref{fig: transport} (b), $\bar{J}_x$ is indeed dependent on the magnetic flux. This originates from the magnetic flux dependence of localization of the quantum bridge states.  

One could generalize above argument related to the sub-dimensional transport for the case of more number of bridges. Note that the number of bridges depend on the translation of the four-dimensional hypercubic lattice. We demonstrate the quantum transports with two and four bridges in Supplemental Materials. As the number of bridges increases, the wave function can be transmitted in at most eight different directions. We would suggest the angle-dependent current measurement controlled by magnetic flux along the bridges.

\section{Conclusion}
\label{sec: conclusion}
In summary, we discussed the new quantum phenomena relevant to the exotic tiling pattern in quasicrystals. Particular translations in four dimensional hypercubic lattice, give rise to the atomic rearrangement where two distinct phason flips coexist and thus result in new local tiling patterns. In such system, we discovered the sub-dimensionally localized quantum bridge states and their relevant transports. 
In the presence of particular magnetic field, 
the wave function tends to be more strongly localized on the bridges forming the Aharonov-Bohm like cages.
It allows the electron jumps from one site to others that share the same local pattern with time evolution. Interestingly, the amount of current can be also controlled in terms of magnetic flux. Furthermore, the number of bridges which are the paths of the sub-dimensional transport, depends on the four-dimensional hypercubic lattice translations. Our study paves a way for finding  unique quantum transports and their magnetic field control in quasicrystals with distinct local tiling patterns.



\section*{Acknowledgements}
This work is supported by National Research Foundation Grant (NRF-2020R1A4A3079707, NRF grand 2021R1A2C1093060).

\begin{appendix}



\end{appendix}



\bibliography{reference}

\newpage

\begin{widetext}
\setcounter{equation}{0}
\setcounter{figure}{0}
\setcounter{table}{0}
\setcounter{page}{1}
\renewcommand{\theequation}{S\arabic{equation}}
\renewcommand{\thefigure}{S\arabic{figure}}
\renewcommand{\theHfigure}{S\arabic{figure}}
\renewcommand{\bibnumfmt}[1]{[S#1]}
\renewcommand{\citenumfont}[1]{S#1}
\renewcommand{\citenumfont}[1]{\textit{#1}}

\begin{center}
\textbf{\large Supplementary Materials of \textit{Quantum bridge states} and \\ sub-dimensional transports in quasicrystals
}
\end{center}

\author{Junmo Jeon, SungBin Lee}


\normalsize

\mdseries

\maketitle


\section{Review: New tiling patterns --- \textit{bridges}}
\label{sec: review}
Here, we briefly review new tiling pattern, say bridges pattern derived from some higher-dimensional translation in the cut-and-project scheme. Starting from a standard example of the quasicrystal, the Ammann-Beenker tiling descended from four-dimensional hypercubic lattice and its projection, we have found that the simple four-dimensional hypercubic lattice shifts could alter the order of rotational symmetry, and create a new quasicrystalline tiling pattern.

Let us briefly review the cut-and-project scheme (CPS). The CPS is a way to understand non-crystallographic symmetry of the low-dimensional quasicrystals in terms of the  high-dimensional lattices\cite{senechal1996quasicrystals}.
Let high-dimensional cubic lattice $\mathcal{L}=\{ \boldsymbol{x} | \boldsymbol{x}=m_i \boldsymbol{e}_i, m_i\in\mathbb{Z}, 1\le i\le D\}$, where $D$ is the dimension of the lattice and $\boldsymbol{e}_i$ is a standard unit vector. The $D$-dimensional space is decomposed into two projection maps $\boldsymbol{\pi}$ and $\boldsymbol{\pi}^{\perp}$. Each of them projects the lattice points in $\mathcal{L}$ onto the subspace of the quasicrystal(QC) (physical space) and its orthogonal complement subspace (perpendicular space), respectively. 
 To produce the nontrivial quasicrystalline pattern, the physical space should have an irrational angle to the lattice surface. 
However, for such an irrational angle, the images of the projection of entire lattice points, $\boldsymbol{\pi}(\mathcal{L})$, densely cover the physical space. 
Thus, one should select a subset of $\mathcal{L}$, which is the compact subset of the perpendicular space $\boldsymbol{\pi}^{\perp}$ that is closed and bounded. This compact subset is often termed as the window, say $K$.
 Now, we project the lattice points only when those image of $\boldsymbol{\pi}^{\perp}$ belongs to the window $K$. Then, the resultant projection image onto the physical space emerges the discrete quasicrystalline lattice structure. As a standard choice of the window, $K=\boldsymbol{\pi}^{\perp}(\mathcal{W}(0))$, where, $\mathcal{W}(0)$ is the Wigner-Seitz cell of the origin\cite{senechal1996quasicrystals,cryst8100370,cryst7100304}.

To illustrate the details, let us consider the example of the CPS, the Ammann-Beenker tiling. The Ammann-Beenker tiling is generated by the CPS with four-dimensional hyper-cubic lattice with $D=4$. The Wigner-Seitz cell is given by the polytope whose 16 verticies are $(\pm\frac{1}{2},\pm\frac{1}{2},\pm\frac{1}{2},\pm\frac{1}{2})$ where the signs of each component are independent. 
The $\boldsymbol{\pi}$ and $\boldsymbol{\pi}^{\perp}$ projections are given by,
\begin{align}
\label{projection}
\boldsymbol{\pi} &=\begin{pmatrix} 1 & \frac{1}{\sqrt{2}} & 0 & -\frac{1}{\sqrt{2}} \\ 0 & \frac{1}{\sqrt{2}} & 1 & \frac{1}{\sqrt{2}} \end{pmatrix} \\
\label{projectionperp}
\boldsymbol{\pi}^{\perp}&=\begin{pmatrix} 1 & -\frac{1}{\sqrt{2}} & 0 & \frac{1}{\sqrt{2}} \\ 0 & \frac{1}{\sqrt{2}} & -1 & \frac{1}{\sqrt{2}} \end{pmatrix}.
\end{align}
Then, the window $K$ becomes a tesseract graph whose vertices are $\boldsymbol{\pi}^{\perp}$-projection images of the vertices of the Wigner-Seitz cell (See Fig.\ref{fig: windowsupp} (a).)\cite{baake2017aperiodic}. The Ammann-Beenker tiling (See Fig.\ref{fig: windowsupp} (c)) is given by the $\boldsymbol{\pi}$-projection of four-dimensional hyper-cubic lattice points, whose image of $\boldsymbol{\pi}^{\perp}$ belongs to the window $K$ in Fig.\ref{fig: windowsupp} (a).
\begin{figure}[]
\centering
\includegraphics[width=0.8\textwidth]{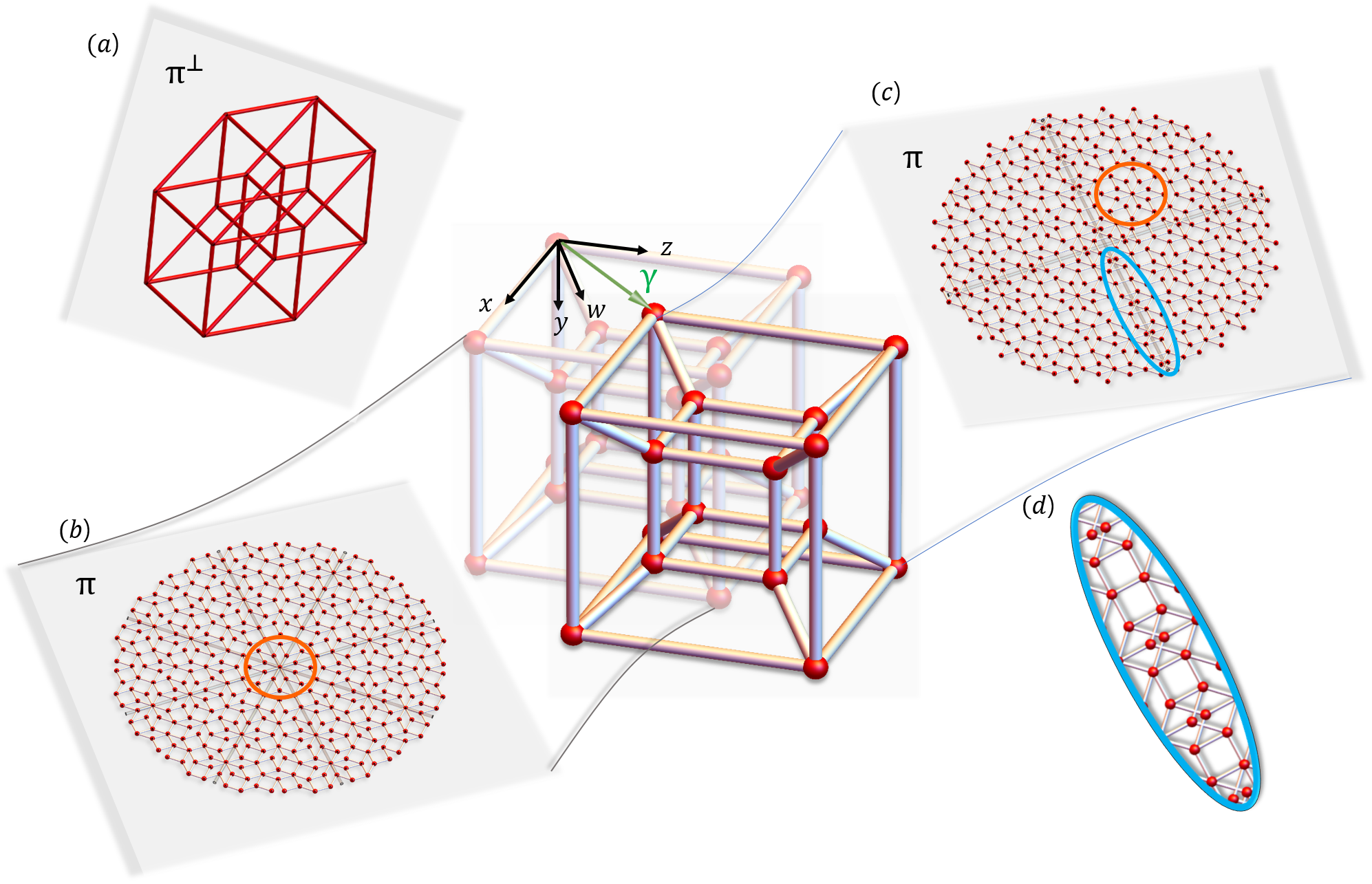}
\caption{\label{fig: windowsupp} Schematic illustration of formation of the quasicrystals with coexisting phason flips from a translation operation in the high-dimensional lattice. The unit cell of four-dimensional hypercubic lattice -- a tesseract, is drawn at the center. 
(The faint image in the back is a tesseract before translation with a shifting vector $\gamma$.) (a) The compact window formed by the $\boldsymbol{\pi}^\perp$-projection of the Wigner-Seitz cell.
The special set of translations given by the shift vector $\boldsymbol{\gamma}$ (green arrow) of the high-dimensional lattice generates quasicrystals with distinct tiling pattern with coexisting atomic positions which are related by the phason flips. Before translation, the $\boldsymbol{\pi}$-projected quasicrystal is the 8-fold symmetric Ammann-Beenker tiling. (c) With translation by  $\boldsymbol{\gamma}=(x,y,z,w)=\frac{1}{2}(1,0,1,0)$, the $\boldsymbol{\pi}$-projection results in a 4-fold symmetric QC with new local tiling pattern. 
In both (b) and (c), the local 8-fold symmetry is shown as marked in orange circle. 
(d) Enlarged image of the new sub-dimensional tiling pattern shown as a blue circle in (c), which is absent in the pure Ammann-Beenker tiling (b).
}
\end{figure}

In general, the higher-dimensional lattice translation in CPS gives unique degrees of freedom of quasicrystalline structure that is the rearrangement of atoms called by the phason flips. According to the phason flips, the tiling pattern of the quasicrystal could be locally deformed. Two or more atomic arrangement would locally compete depending on the higher-dimensional lattice translation as the phason flips. For the case that one of the competing atomic arrangement is favored, the kind of local tiling patterns would not be drastically changed, and hence the tilings are locally isomorphic. However, for some special higher-dimensional lattice shifts, the locally competing atomic arrangements would be coexisting, thus, the distinct local tiling patterns would be appeared.

As a specific example, let us consider the CPS for the Ammann-Beenker tiling illustrated in Fig.\ref{fig: windowsupp} again. Generally, for a hypercubic lattice shift, $\boldsymbol{\gamma}$, the tiling pattern could be locally altered by selecting one of two competing local atomic arrangements. However, for a special $\boldsymbol{\gamma}=\frac{1}{2}(1,0,1,0)$, the compact (closed and bounded) window accepts both competing local atomic positions related by the phason flilp. Fig\ref{fig: windowsupp} (d) shows the local tiling pattern comprised of many such coexisting atomic arrangements related by the phason flips. It turns out that the distinct quasicrystalline tiling pattern in Fig.\ref{fig: windowsupp} (c) from the pure Ammann-Beenker tiling is formed at such special hypercubic lattice translation $\boldsymbol{\gamma}$.

The new tiling pattern called by the bridge illustrated in Fig.\ref{fig: windowsupp} (d) is sub-dimensionally extending. The directions of the extension of the bridges are depending on the hypercubic lattice shift $\boldsymbol{\gamma}$, and they could be 1,2, and 4 different directions depending on the $\boldsymbol{\gamma}$\cite{jeon2021discovery}. See Sec.\ref{sec: 3} for the demonstration of the sub-dimensional quantum transport along the bridges with two and four different directions.

Although the randomized phason flips just give defect, here, the coexisting phason flips appear in a quasi-periodically ordered way. Hence, the tiling pattern in Fig.\ref{fig: windowsupp} (d) is ordered quasicrystalline structure rather than random defects.

\section{Irregular diffusion on the bulk}
\label{sec: 2}
For comparison, we investigate a typical dynamics in the new quasicrystalline tiling starting from the initial $\delta$-function-like wave function placed on a random bulk site out of the bridge pattern (bllue shaded region in Fig.\ref{fig: diffusion}). In detail, Fig \ref{fig: diffusion} illustrates the probability distribution (colored) of the wave function evolved from the initial position $X_0$ out of the bridge pattern after time $\tau=3000t^{-1}$. Note that the quasicrystal has the bridge pattern along the $x$-axis (blue shaded region). Fig.\ref{fig: diffusion} demonstrates that the dynamics of the wave function becomes the irregular diffusion in the bulk if the initial position is out of the bridges pattern. Here, the flux is given by $1.35\phi_0$ where $\phi_0$ is the flux quanta. Hence, we can conclude that the electronic transport along the bridges discussed in the main text is indeed an unique phenomenon dependent on the initial position.
\begin{figure}[]
\centering
\includegraphics[width=0.7\textwidth]{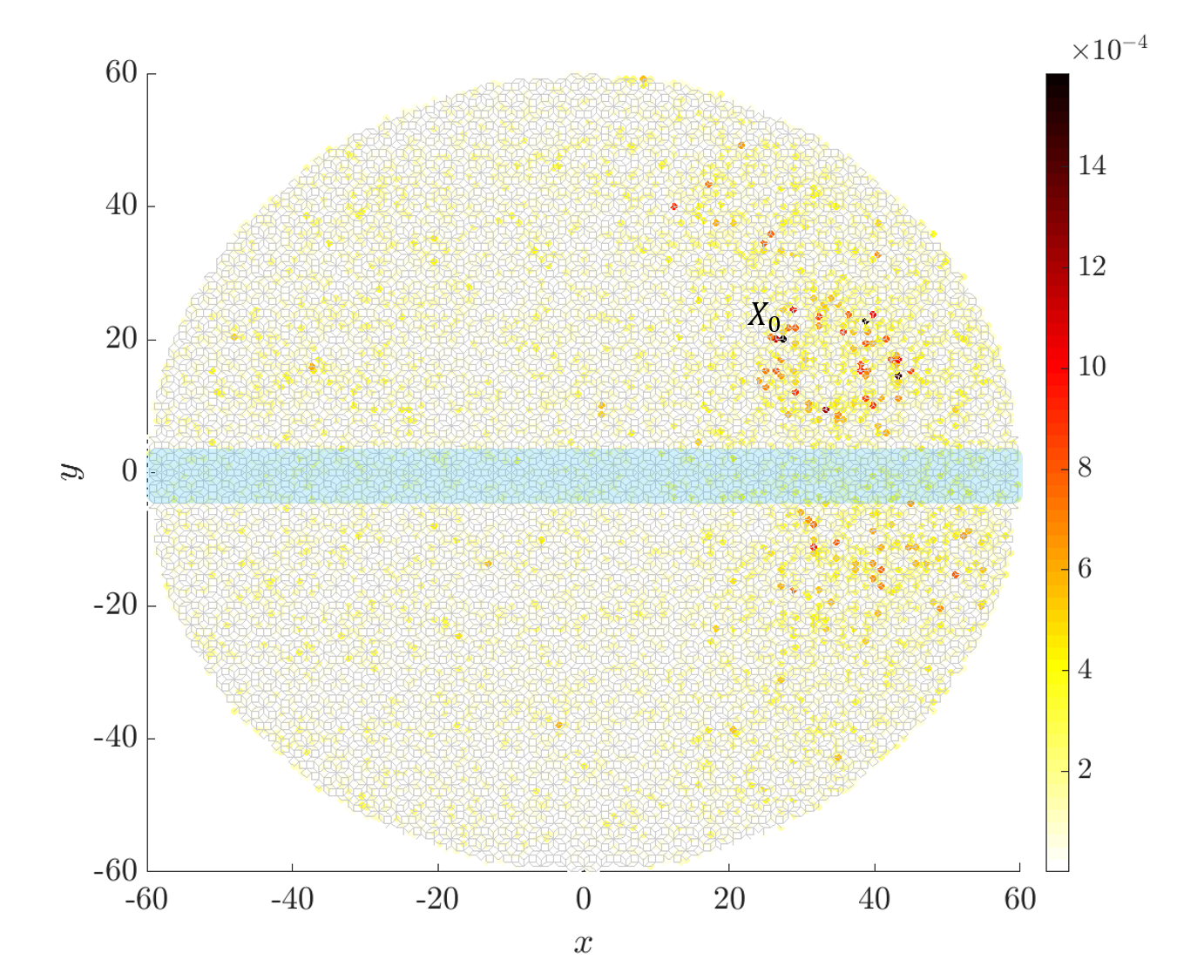}
\caption{\label{fig: diffusion} Time evolution after time $\tau=3000 t^{-1}$ of the $\delta$-function-like wave function initially placed on general bulk site, $X_0$ out of the bridges represented by the blue shaded region. Here, $t$ is the hopping magnitude in the tight-binding Hamiltonian. See the main text for detailed infromation of the Hamiltonian. The magnetic flux is $1.35\phi_0$. The quasicrystal is two-fold symmetric with the bridges pattern along $x$-axis.}
\end{figure}


\section{Transports with more number of the bridges}
\label{sec: 3}
The number of new tiling pattern, bridges could be two and four depending on the hypercubic lattice shifts\cite{jeon2021discovery}. Thus, for completeness of our discussion, we illustrate similar quantum transports with two and four numbers of bridges. Figs.\ref{fig: more} (a,b) and (c,d) demonstrate the quantum transports along with the two and four numbers of bridges, respectively. The $A$ in Figs.\ref{fig: more} (a,c) indicate the initial positions whose local patterns are the same as the local pattern of the $A$ site discussed in the main text.

To be more specific, Figs.\ref{fig: more} (a,c) and (b,d) show the probability distributions after time $\tau=1000 t^{-1}$ and $\tau=3000 t^{-1}$, respectively.  Here, the fluxes are set to be $1.35\phi_0$ where $\phi_0$ is the flux quanta. Note that the quantum transports are both sub-dimensional along the bridges patterns. Importantly, as the numbers of the bridges pattern increases, the sub-dimensional wave function transports can occur in various directions. It turns out that the angle-dependent sub-dimensional transports can emerge with different numbers of bridges.
\begin{figure}[]
\centering
\includegraphics[width=0.9\textwidth]{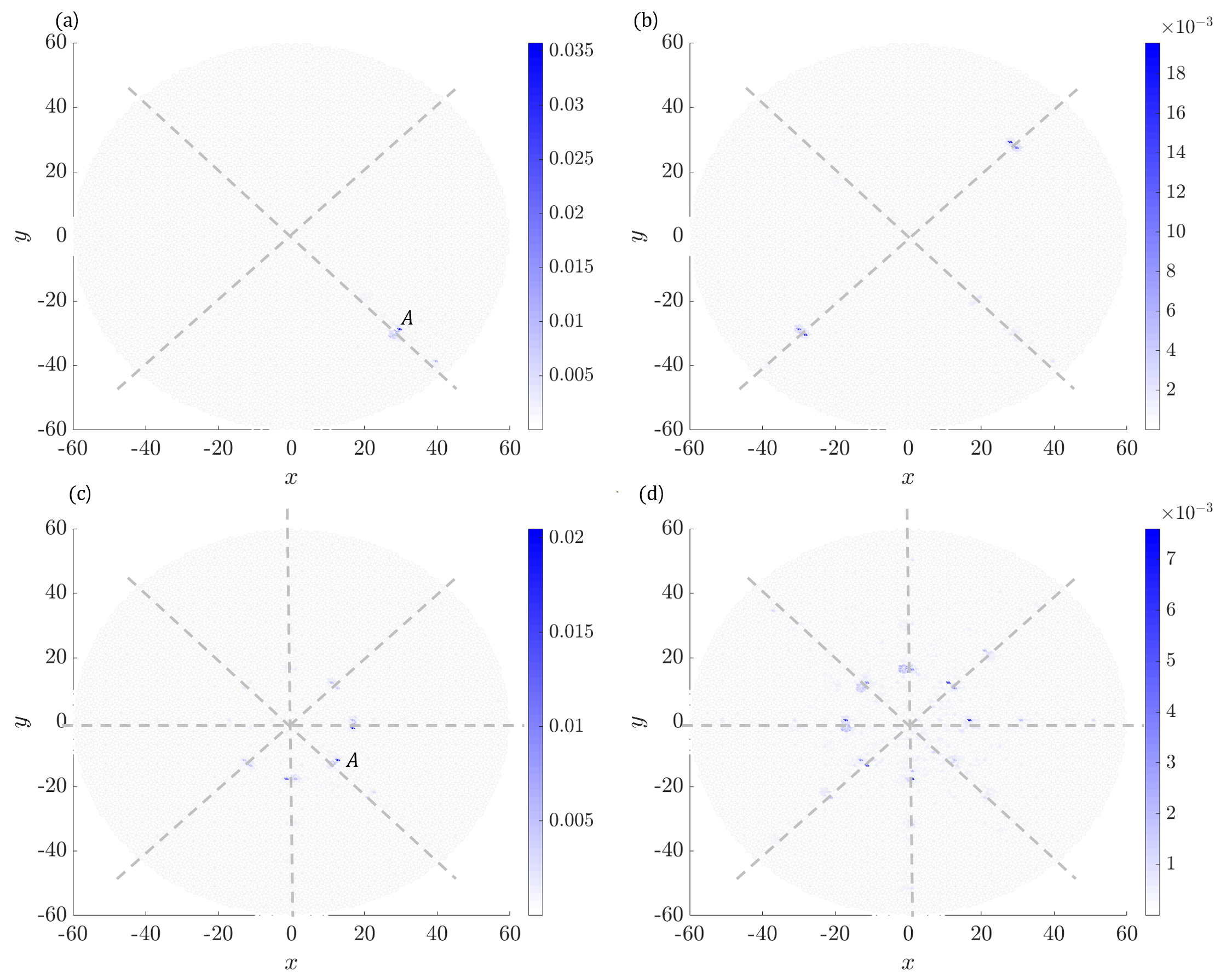}
\caption{\label{fig: more} New quasicrystals descended from the four-dimensional lattice translations (a,b) $1/2(1,0.1,0)$ and (c,d) $1/2(1,1,1,1)$ where the lattice constant of the hypercubic lattice is unity. Each new quasicrystal has four-fold (a,b) and eight-fold (c,d) symmetric bridges patterns. Dashed lines are drawn to emphasize the bridges tiling pattern region. Under the flux $\Phi=1.35\phi_0$, (a,c) and (b,d) show the probability distributions after time $\tau=1000 t^{-1}$ and $\tau=3000 t^{-1}$, respectively. The initial positions of the $\delta$-function-like wave function are represented by $A$ in (a,c) whose local patterns are the same as the local pattern of the $A$-site discussed in the main text.}
\end{figure}
\end{widetext}

\begin{appendix}



\end{appendix}




\end{document}